\documentclass[10pt, conference]{IEEEtran}
\IEEEoverridecommandlockouts
\usepackage[noadjust]{cite}
\usepackage{amsmath, amssymb, amsfonts, url, amsthm}
\usepackage[ruled,vlined]{algorithm2e}
\usepackage{graphicx, tabularx, booktabs, subcaption}
\usepackage{textcomp}
\usepackage{xcolor}
\usepackage{makecell}
\usepackage[acronym]{glossaries}

\newtheorem{lemma}{Lemma}
\newtheorem{theorem}{Theorem}

\newtheorem{remark}{Remark}

\DeclareMathOperator*{\argmin}{arg\min}
\DeclareMathOperator*{\argmax}{arg\max}
\def\BibTeX{{\rm B\kern-.05em{\sc i\kern-.025em b}\kern-.08em
    T\kern-.1667em\lower.7ex\hbox{E}\kern-.125emX}}

\begin{document}

\title{Type-based Over-the-Air Computation
\thanks{This work is part of the project SOFIA PID2023-147305OB-C32 funded by MICIU/AEI/10.13039/501100011033 and FEDER/UE.}
}

\author{\IEEEauthorblockN{Marc Martinez-Gost\IEEEauthorrefmark{1}, Ana Pérez-Neira\IEEEauthorrefmark{1}\IEEEauthorrefmark{2}}
\IEEEauthorblockA{
\IEEEauthorrefmark{1}Centre Tecnològic de Telecomunicacions de Catalunya, Spain\\
\IEEEauthorrefmark{2}Dept. of Signal Theory and Communications, Universitat Politècnica de Catalunya, Spain\\
mmartinez@cttc.es
}}

\newacronym{AI}{AI}{Artificial Intelligence}
\newacronym{AirComp}{OAC}{over-the-air computation}
\newacronym{AWGN}{AWGN}{additive white Gaussian noise}
\newacronym{CNN}{CNN}{convolutional neural network}
\newacronym{CSI}{CSI}{channel state information}
\newacronym{DA}{DA}{direct aggregation}
\newacronym{DSB}{DSB}{double sideband}
\newacronym{FL}{FEEL}{federated learning}
\newacronym{FSK}{FSK}{frequency shift keying}
\newacronym{MSE}{MSE}{mean squared error}
\newacronym{NMSE}{NMSE}{normalized mean squared error}
\newacronym{PAM}{PAM}{pulse amplitude modulation}
\newacronym{PPM}{PPM}{pulse position modulation}
\newacronym{TBMA}{TBMA}{type-based multiple access}
\newacronym{SNR}{SNR}{signal-to-noise ratio}

\maketitle
\begin{abstract}
Over-the-air computation (OAC) exploits the superposition property of wireless channels to compute functions of distributed data, reducing the communication overhead associated with its multiple access. However, existing OAC frameworks are often designed around specific classes of functions, limiting their applicability to more general nonlinear computations. This paper investigates type-based multiple access (TBMA) as a general framework for OAC, enabling the receiver to recover a histogram of the distributed observations and compute a broad class of functions without modifying the underlying communication scheme.
We formulate TBMA as a histogram estimation problem and propose two digital detectors. We prove that a low-complexity successive cancellation algorithm achieves the optimal maximum-likelihood (ML) performance. We derive analytical expressions for the mean squared error (MSE) and compare TBMA with conventional schemes in terms of computation accuracy, transmission duration, bandwidth and energy consumption. Numerical experiments validate the analytical results and demonstrate that, unlike existing OAC techniques, TBMA exhibits an exponential MSE decay with increasing signal-to-noise ratio (SNR) and constant transmit energy, independently of the function being computed. These findings position TBMA as a step toward more versatile OAC architectures, compatible with existing transmission schemes, and broadening the range of functions that can be computed directly over the wireless channel and enabling new possibilities for distributed computation and decision-making.
\end{abstract}


 \section{Introduction}
Conventional wireless communication systems are primarily designed to reliably convey information from transmitters to receivers, typically treating the wireless channel as a source of distortion that must be compensated for. However, the linearity and superposition properties of electromagnetic waves offer an alternative perspective, in which the propagation medium itself can be exploited to perform computation. This principle underlies over-the-air computation (OAC), a communication paradigm in which multiple devices simultaneously transmit their data so that the wireless channel naturally combines their signals, enabling the receiver to directly estimate a desired function of the distributed information without individually decoding each transmission \cite{sp_magazine}. Unlike other multiple-access schemes, which aim to recover individual messages, OAC exploits electromagnetic superposition as a computational resource, potentially improving spectral efficiency and reducing communication latency in applications such as distributed sensing, wireless edge learning and large-scale Internet of Things networks.

A natural realization of this paradigm is linear OAC, in which information linearly modulates the amplitude of the transmitted carrier waveform. By appropriately compensating for channel effects, linear OAC enables the efficient computation at the receiver of linear functions, such as sums and weighted averages, through the coherent superposition of transmitted waveforms. Extensive research has addressed its fundamental challenges, including optimal power allocation \cite{power_control}, robustness to synchronization errors \cite{robust_sync}, and extensions to multi-antenna settings \cite{oac_mimo}. A comprehensive overview of linear OAC can be found in \cite{survey}. However, extending this approach to computing more general nonlinear functions remains considerably more challenging. Although nomographic representations can enable the computation of certain nonlinear functions through suitable pre-processing and post-processing, and digital modulation schemes offer alternative implementations, these solutions often depend strongly on the target function. This limitation of existing approaches for nonlinear function computation provides the motivation for investigating alternative multiple-access strategies.

More recently, \gls{TBMA} has emerged as an alternative approach to linear OAC \cite{tbma_fsk, sahin_fsk, martinez2023lora}. TBMA employs $N$ orthogonal accesses that enable the receiver to reconstruct a histogram of the transmitted data. While TBMA may appear to offer lower resource efficiency, the histogram represents a useful tool that can be exploited for certain applications, such as noncoherent aggregation for federated learning \cite{sahin_fsk}, client selection in heterogeneous federated learning \cite{ngo-feel}, robust aggregation against Byzantine attacks \cite{tbma_robust}, distributed data processing with Map-Reduce \cite{mapreduce}, outage mapping for emergency communications \cite{emergency}, consensus algorithms \cite{consensus} and unsourced multiple access \cite{tbma-unsourced}. Moreover, TBMA is compatible with general-purpose digital modulations, regardless of the function to be computed.

Rather than viewing TBMA merely as an approximation mechanism for computing functions, it can be understood as a more general framework for conveying information about distributed data \cite{spectrum_apn}. The histogram provides a generic and rich representation that is not tied to any particular function, but can instead support a broad range of tasks, from computing specific quantities to interpreting data and making complex decisions. This perspective fundamentally broadens the scope of OAC beyond function computation toward general task-oriented communication, where the ability to convey and interpret aggregate information can be more valuable than reconstructing individual measurements.

In this work, we provide a systematic comparison of type-based OAC with linear OAC and orthogonal multiple access schemes in terms of computation accuracy and resource efficiency. 
This is the first work to address optimal detection in type-based OAC and to propose a low-complexity algorithm that achieves optimality in the maximum-likelihood (ML) sense \cite{martinez2026tbma_ml}. Our results show that TBMA provides a general framework for OAC that circumvents several limitations of linear OAC, including compatibility with existing communication schemes, fixed energy consumption and exponential error decay with respect to the signal-to-noise ratio (SNR) for arbitrary functions.

\subsection{Motivation}
Early work on OAC demonstrated that uncoded linear analog transmission can achieve optimal mean squared error (MSE) scaling under certain channel models \cite{ota_original}. This approach leverages the natural aggregation performed by the wireless channel, avoiding the need to decode individual transmissions and outperforming conventional communication schemes based on separate source and channel coding. However, conventional analog OAC is primarily suited to the computation of linear functions.

One approach to overcoming this limitation relies on nomographic decompositions \cite{goldenbaum}. This framework enables the computation of a class of nonlinear functions by decomposing the target function into distributed preprocessing at each transmitter, summation over the multiple-access channel and postprocessing at the receiver. Although mathematically general, this approach can introduce significant practical challenges when applied to communication signals. Consider, for example, the product function, which can be decomposed as
\begin{equation}
    \prod_{k=1}^K x_k=\exp\left(
    \sum_{k=1}^K \ln(x_k)
    \right),\quad x_k>0.
    \label{eq:prod}
\end{equation}

Considering a baseband model, each transmitter applies logarithmic pre-processing to its observation $x_k$ before transmission, and the receiver applies an exponential transformation to the aggregated signal. On one hand, the logarithmic pre-processing can substantially increase the dynamic range of the transmitted signals and potentially lead to excessive transmit power. On the other hand, in the presence of additive channel noise, the exponential post-processing transforms the noise statistics and can produce signal-dependent, non-Gaussian errors. This transformation complicates receiver design, as conventional components such as symbol detectors, synchronization units and timing recovery modules are often designed under additive white Gaussian noise (AWGN) assumptions. Departures from these assumptions can degrade their performance and necessitate more sophisticated receiver processing or higher SNR \cite{log-fsk}. Beyond these implementation challenges, nomographic decompositions are not universally applicable. For example, the logarithmic decomposition in \eqref{eq:prod} is valid only for strictly positive inputs. This is, the existence of a suitable nomographic decomposition is not guaranteed for arbitrary nonlinear functions, limiting the generality of this approach to OAC.

An alternative approach is to implement OAC using function-specific nonlinear digital modulations \cite{channelcomp}. The transmitted symbols are jointly designed according to the target function so that their superposition at the receiver produces a composite constellation in which different function values can be distinguished reliably. Although this original strategy directly accommodates nonlinear computation, its flexibility comes at the cost of increased design complexity. In particular, the constellation depends on the target function, the number of possible function values and the number of transmitters. Furthermore, solving the associated optimization problem becomes increasingly difficult as the number of transmitters and constellation symbols (i.e., function values) grows. These dependencies limit the scalability of the approach and complicate its extension to general functions and high-resolution computation.

TBMA has recently emerged as an alternative framework for OAC. Unlike the previous approaches, it uses generic orthogonal digital modulations whose superposition enables the receiver to recover a histogram of the distributed observations. The original TBMA framework was developed for distribution-based parametric estimation \cite{tbma_tong} (e.g., estimating the probability of a Bernoulli distribution). The proposed estimator was shown to asymptotically approach the ML estimator and can be formulated as the minimization of the Pearson chi-square test. This criterion quantifies the difference between the empirical distribution obtained from the observations and the distribution predicted by an assumed statistical model.

Despite its asymptotic optimality, the original TBMA estimation framework presents several limitations when applied to general OAC problems:
\begin{enumerate}
    \item The estimator requires prior knowledge of the underlying data distribution. This assumption can be restrictive in distributed sensing and learning applications, where the statistical properties of the observations may be unknown or vary over time.
    
    \item The estimator is derived under an asymptotic regime in which the number of transmitters tends to infinity. With a finite number of devices, the empirical histogram may provide an imperfect representation of the distribution, potentially affecting estimation accuracy. Moreover, when the number of transmitters is relatively small, the objective of OAC is to recover the exact aggregate function rather than estimate a statistical parameter or expectation. Therefore, a parametric estimator may not be the most appropriate solution.
    
    \item The original formulation assumes that the observations follow a discrete distribution. Although sufficiently fine quantization can approximate continuous-valued observations with limited quantization distortion, coarse quantization may introduce substantial errors. Depending on the quantization scheme, these errors can be data-dependent and alter the empirical distribution used by the estimator. Furthermore, the original formulation does not directly address the computational complexity of optimization over a continuous parameter space.
\end{enumerate}

In this work, we propose TBMA as a general framework for OAC. Our approach builds on the observation that any symmetric function of the observations is fully determined by their histogram \cite{sahin2026generic}. Therefore, TBMA allows to separate the communication problem (i.e., histogram estimation) from the computation problem (i.e., function evaluation). This separation allows histogram estimation to be approached as a conventional digital communication problem, with a transmission scheme that remains fixed regardless of the target function.

\subsection{Contributions}

OAC is valuable when the reduction in the required communication resources obtained through analog superposition produces a meaningful end-to-end advantage over conventional digital communication and computation, once channel estimation, synchronization, power control, noise and application-level accuracy are included. The overarching goal of this paper is to show the potential of TBMA  as a general framework for OAC. The main original contributions of this work are summarized as follows:

\begin{enumerate}
    \item We formulate a communication framework for OAC based on TBMA and we propose two histogram detection strategies: independent bin detection and joint ML detection. We characterize both their symbol error probability (SER) in histogram detection and their mean squared error (MSE) in function computation. The ML detection is shown to admit an optimal low-complexity greedy implementation.

    \item We conduct a systematic comparison of TBMA with conventional schemes, namely orthogonal transmission schemes such as time division multiple access (TDMA), and linear OAC, for the computation of the sample mean. For each scheme, we derive analytical expressions for the MSE in function computation, transmission duration, bandwidth requirements and energy consumption.

    \item We validate the theoretical analysis through numerical evaluations and simulations, demonstrating the potential advantages of TBMA-based OAC for computing nonlinear functions.
\end{enumerate}

\subsection{Paper Organization and Notation}
The rest of this paper is organized as follows. The remainder of Section I reviews the related literature and summarizes the main contributions of this paper. Section II introduces the system model and section III reviews conventional communication approaches for function computation. Section IV introduces TBMA, presents several detection strategies and provides a systematic comparison. Section V complements it with numerical results and Section VI concludes the paper.

Throughout this paper, scalars are denoted by italic letters and vectors by boldface letters. The $n$-th entry of vector $\mathbf{s}_k$ is denoted by \(s_{k,n}\). The sets of real numbers and non-negative integers are denoted by $\mathbb{R}$ and $\mathbb{Z}_+$, respectively. Finally, \mbox{$x\sim\mathcal{U}(a,b)$} denotes a random variable uniformly distributed over the interval $[a,b]$.

\section{System Model}
Consider a wireless network consisting of $K$ distributed devices (transmitters) and a single server (receiver). Each device $k$ has a local data symbol $d_k$ (e.g., sensor measurements or the result of a local computation), where the data symbols are assumed to be independent and identically distributed (i.i.d.). The data symbols are bounded as \mbox{$-A< d_k < A$} with variance $\mathbb{E}\big\{|d_k|^2\big\}=\alpha A^2$, where $0<\alpha\leq 1$ depends on the specific distribution.

Each transmitter generates a communication symbol \mbox{$\mathbf{s}_k\in\mathbb{R}^N$} from $d_k$, where $N$ and the specific mapping depend on the modulation format. Then, it encodes $\mathbf{s}_k$ into a continuous-time baseband waveform $x_k(t)$ over a symbol interval $T$, 
\begin{equation}
    x_k(t)=\sum_{n=1}^N s_{k,n}\varphi_n(t),
    \quad
    0\leq t<T,
\end{equation}
where $\{\varphi_1^{}(t),\ldots,\varphi^{}_N(t)\}$ is a set of orthonormal basis functions, this is,
\begin{equation}
    \int_{-\infty}^\infty \varphi_i^{}(t)\varphi_j^*(t),dt=\delta_{ij}
    \label{eq:ortho}
\end{equation}

All devices transmit simultaneously and perfectly synchronized. Assuming ideal channel compensation, the received signal is
\begin{equation}
    y(t)=\sum_{k=1}^K x_k(t)+w(t),\quad
    0\leq t<T,
\end{equation}
where $w(t)$ denotes additive white Gaussian noise (AWGN). The received signal $y(t)$ is processed to ultimately estimate the desired function ${f}(d_1,\dots,d_K)$.


Currently, OAC is restricted to symmetric functions, i.e., functions whose output is invariant to the ordering of the input data. We first consider the traditional sample mean function,
\begin{equation}
    f(d_1,\dots,d_K)=\frac{1}{K}\sum_{k=1}^K d_k,
    \label{eq:mean_function}
\end{equation}
and in section \ref{sec:results} we discuss the extension to nonlinear functions. Note that although $d_k$ might be modeled with underlying distribution, equation \eqref{eq:mean_function} does not correspond to the statistical mean of the distribution. For instance, with $K=1$ user, the function returns $d_1$ exactly.

The figure of merit considered throughout this work is the mean squared error (MSE),
\begin{equation}
\text{MSE} = \mathbb{E}\bigg\{\left(f(d_1,\dots,d_K)-\hat{f}(d_1,\dots,d_K)\right)^2\bigg\},
\label{eq:figure_merit_mse}
\end{equation}
where $\hat{f}$ is the function estimated at the receiver side and the expectation is taken over the data and channel noise. Throughout this work we will also asses other relevant metrics in communications, namely duration time, bandwidth and energy consumption.

\subsection{Channel and Synchronization Assumptions}

In this paper, we assume that all devices transmit simultaneously with ideal time and phase synchronization. This assumption is commonly adopted in the theoretical analysis of OAC, as it allows us to isolate the fundamental performance of the considered schemes from synchronization impairments. In practice, synchronization can be achieved using appropriate techniques \cite{sahin_sync, dahl2026decentralized}. Furthermore, we assume that each device performs perfect channel inversion \cite{power_control}, which requires channel state information at the transmitters and enables coherent signal superposition. Since the receiver observes an aggregate signal, individual channel coefficients cannot generally be compensated for through conventional post-processing alone. Alternatively, recent advances in electromagnetic waveform processing offer additional opportunities to engineer the propagation environment through reconfigurable scattering devices, facilitating the desired signal aggregation \cite{em_processing}. Both approaches exploit the linearity and superposition properties of electromagnetic fields.

To the best of our knowledge, this is the first work to investigate TBMA in the context of OAC and systematically compare it with other OAC approaches. Accordingly, we adopt idealized channel and synchronization assumptions to establish a common reference framework for a fair comparison and to isolate the fundamental performance differences between the considered schemes. These assumptions provide a benchmark against which practical impairments can be assessed. Extensions to noncoherent transmission, imperfect channel compensation and synchronization errors are left for future work.

\subsection{Impact of Quantization}
Most of the schemes considered in this work are digital and therefore involve quantization of the data prior to transmission. Even in the absence of channel errors, the receiver can only recover the quantized representation of the data, which introduces distortion in the final function estimate. To characterize this effect separately from channel-induced distortion, we first quantify the contribution of data quantization to the overall MSE.

Without loss of generality we consider a uniform mid-rise quantizer with $N$ levels that maps $d_k$ into one integer:
\begin{equation}
    \bar{d}_k=\mathcal{Q}(d_k)=
    \biggl \lfloor
    N\frac{d_k+A}{2A}
    \biggl \rfloor +1\in\{1,\dots,N\}
\end{equation}

At the receiver, the data are reconstructed by applying the inverse quantization mapping to the received index:
\begin{equation}
    \tilde{d}_k=\mathcal{Q}^{-1}(\bar{d}_k)=-A +\frac{2A}{N} \left(\bar{d}_k-\frac{1}{2}\right)
\end{equation}

To compute the MSE, we first define
\begin{equation}
    \tilde{f}=\frac{1}{K}\sum_{k=1}^K \tilde{d}_k
\end{equation}
as the sample mean computed from the quantized data. The MSE can then be decomposed as
\begin{align}
    \text{MSE}&=
    \mathbb{E}\Big\{\left(\left(f-\tilde{f}\,\right)+\left(\tilde{f}-\hat{f}\right)\right)^2\Big\}\nonumber\\
    &=\underbrace{\mathbb{E}\Big\{\left(f-\tilde{f}\,\right)^2\Big\}}_{\text{MSE}_\text{q}}+
    \underbrace{\mathbb{E}\Big\{\left(\tilde{f}-\hat{f}\right)^2\Big\}}_{\text{MSE}_\text{c}},
    \label{eq:independence_mse}
\end{align}
where $\text{MSE}_\text{q}$ correspond to the quantization distortion and $\text{MSE}_\text{c}$ to channel-induced distortion. The last equality follows from the assumption that both errors are uncorrelated. 

For high-resolution quantization (i.e., large $N$), the quantization error can be modeled as an independent uniformly distributed random variable over one quantization interval. This results in
\begin{align}
    \text{MSE}_\text{q}&= \mathbb{E}\Bigg\{\left(\frac{1}{K}\sum_{k=1}^K d_k-\frac{1}{K}\sum_{k=1}^K\tilde{d}_k\right)^2\Bigg\}\nonumber\\
    &=\frac{1}{K}\mathbb{E}\Big\{ \left(d_k-\tilde{d}_k\right)^2\Big\}\nonumber\\
    &=\frac{A^2}{3N^2K}
    \label{eq:mse_q}
\end{align}

This quantization error represents a floor on the MSE that persists at high SNR, when the effect of channel noise becomes negligible.
The remainder of this paper is devoted to deriving the channel-induced MSE component for the different communication schemes.

\section{Conventional schemes}
This section reviews conventional communication schemes considered in the state of the art that can serve as baselines. Specifically, we consider orthogonal multiple access 
and linear OAC. For each scheme, we derive the corresponding theoretical performance metrics, which provide the basis for motivating the introduction of TBMA in the following section.

\subsection{Orthogonal Multiple Access}
\label{sec:ortho}
From a conventional communication perspective, interference is generally viewed as undesirable. One natural approach is to avoid simultaneous transmissions by allocating orthogonal resources to the different devices, allowing the receiver to recover each transmitted data symbol individually. We consider TDMA\footnote{Other orthogonal resources, such as frequency, code or spatial dimensions, can be used instead, with corresponding trade-offs in resource consumption.} as the conventional communication baseline, where devices transmit sequentially in separate time slots.\newline

\noindent
\textbf{Modulation}: Since each time slot is dedicated to a point-to-point transmission, any modulation format may be adopted. To enable a fair comparison with \gls{TBMA}, we consider a digital implementation in which each device first quantizes its data and then maps the quantized value to a communication symbol. Specifically, each transmitter maps $\bar{d}_k$ to $\mathbf{s}_k$ as 
\begin{equation}
    s_{k,n}=
    \begin{cases}
        1,\quad \bar{d}_k=n\\
        0,\quad \bar{d}_k\neq n.
    \end{cases}
\end{equation}
This modulation format is known in the literature as index modulation, in which each quantized value is represented by a distinct communication dimension \cite{index_modulation}. Throughout this paper we consider $N$-ary FSK, where each data index is mapped to a distinct frequency subcarrier, and other modulation formats are discussed in Section \ref{sec:discussion}.

The resulting symbol is then mapped to a continuous-time baseband waveform as
\begin{equation}
    x_k(t) = 
    \sqrt{P}\sum_{n=1}^N s_{k,n}\varphi_{n}^{}(t)
    =\sqrt{P}\,\varphi_{\bar{d}_k}^{}(t),
    \quad 0\leq t<T
    \label{eq:da_tx}
\end{equation}
\newline

\noindent
\textbf{Duration and bandwidth:} Since users are scheduled with TDMA, the total transmission duration is $D=KT$. In the considered FSK scheme, each user occupies one of the $N$ frequency subcarriers, resulting in an overall bandwidth of $B=N/T$.\newline

\noindent
\textbf{Energy:} The symbol energy per device is obtained from the autocorrelation function,
\begin{align}
    E_s&=\int_{-\infty}^\infty|x_k(t-mT)|^2dt=P\int_{-\infty}^\infty|\varphi_{\bar{d}_k}(t-mT)|^2dt\nonumber\\
    &=P\int_{-\infty}^\infty|\varphi_{\bar{d}_k}(\tau)|^2d\tau=P
\end{align}
where $m$ indexes the data symbol. The third equality is obtained with a change of variables, $\tau=t-mT$, and the latter follows from \eqref{eq:ortho}. As expected, in an orthogonal signaling scheme, the transmitted energy does not depend on the data because the information is modulated in the choice of resource.\newline

\noindent
\textbf{Receiver processing}:
The received signal for transmitter $k$ is
\begin{align}
    y_k(t)&=\sqrt{P}\,\varphi_{\bar{d}_k}^{}(t)+w_k(t),
\end{align}
where $w_k(t)$ is \gls{AWGN} with  power spectral density $N_0/2$. To demodulate the information, the receiver computes a matched filter.
At the output of the $n$-th matched filter, sampled at $t=T$, the signal is $r_{k,n}=\sqrt{P}\delta_{n\bar{d}_k}+\nu_{k,n}$, where $\nu_{k,n}\sim\mathcal{N}(0, N_0/2)$ denotes the noise component.

To recover the data signal, the receiver first takes the largest output,
\begin{equation}
    \hat{{d}}_k=\argmax_n\, |r_{k,n}|,
\end{equation}
whose symbol SER corresponds to the one of orthogonal signaling \cite{proakis2008digital},
\begin{equation}
    P_e=\text{Pr}
    \left(
    \hat{{d}}_k\neq\bar{d}_k
    \right)\approx (N-1)Q\left(\sqrt{\frac{E_s}{N_0}}\right),
\end{equation}
where $Q(\cdot)$ is the Q-function and $E_s/N_0$ is the SNR per user.

The MSE can be computed by applying the law of total expectation, conditioning on whether a detection error occurs:
\begin{align}
    \text{MSE}_{\text{c},k}&=
    \mathbb{E}\bigg\{
    \left(\mathcal{Q}^{-1}(\hat{{d}}_k)-\mathcal{Q}^{-1}(\bar{d}_k)\right)^2\Big|
    \hat{{d}}_k\neq\bar{d}_k
    \bigg\}
    P_e\nonumber\\
    &=
    \frac{4A^2}{N^2}\mathbb{E}\bigg\{
    \left(\hat{{d}}_k-\bar{d}_k\right)^2\Big|
    \hat{{d}}_k\neq\bar{d}_k
    \bigg\}
    P_e
    \nonumber\\
    &=
    \frac{4A^2}{N^2}
    \frac{N(N+1)}{6}
    P_e
    \nonumber\\
    &\approx\frac{2A^2}{3}P_e,
    \label{eq:mse_bound_OMA}
\end{align}
where the  first term corresponds to the MSE introduced by a detection error. Assuming that the decoded index $\hat{d}_k$ is equally likely to be any of the $N-1$ incorrect indices, the conditional MSE is obtained by averaging the squared differences between the transmitted and decoded indices over all possible erroneous decisions. The last approximation follows from large $N$.

This procedure is repeated for all $K$ transmissions and the receiver estimates the sample mean function as
\begin{equation}
    \hat{f}=\frac{1}{K}\sum_{k=1}^K \mathcal{Q}^{-1}(\hat{{d}}_k)
\end{equation}

The overall MSE corresponds to adding the contribution of the $K$ individual transmissions:
\begin{align}
    \text{MSE}_\text{c}= \frac{1}{K}\text{MSE}_{\text{c},k}\approx\frac{2A^2}{3}\frac{N}{K}Q\left(\sqrt{\frac{E_s}{N_0}}\right),
\end{align}
which highlights several relevant trade-offs. First, increasing $N$ reduces the quantization error but increases the channel-induced MSE, as well as the required bandwidth. Second, the MSE decreases as $1/K$, rather than $1/K^2$, since the receiver aggregates $K$ individual estimates and accumulates $K$ independent noise contributions.

\subsection{Linear OAC}
\label{sec:ba}

The traditional approach to OAC is to modulate the information in the amplitude of the transmitted signal.\newline

\noindent
\textbf{Modulation:} The communication symbol is $s_k=d_k$, hence the term linear OAC. The baseband signal transmitted by user $k$ is
\begin{equation}
    x_k(t) = \sqrt{P}\,s_k\varphi(t)=\sqrt{P}\,d_k\varphi(t),
    \quad 0\leq t<T
    \label{eq:da_tx2}
\end{equation}
where $\varphi(t)$ is common waveform shared by all transmitters, and its subscript has been omitted without loss of generality. Equation \eqref{eq:da_tx2} assumes a \gls{DSB} implementation, but a digital $M$-ary \gls{PAM} could be used instead.\newline

\noindent
\textbf{Duration and bandwidth:} All transmitters occupy the same time and frequency resource. Therefore, the overall duration is $D=T$ and the required bandwidth is $B=1/T$.
\newline

\noindent
\textbf{Energy:} The symbol energy per device is
\begin{align}
    E_s&=P|d_k|^2\int_{-\infty}^\infty|\varphi_{}(t-mT)|^2dt=
    P|d_k|^2,
\end{align}
and averaging over the data symbols gives an average symbol energy of $E_s=\alpha A^2P$. Unlike the other schemes considered in this work, the energy of linear modulation depends on the transmitted data. As we will see, this data-dependent energy has a significant impact on the computation of nonlinear functions. \newline

\noindent
\textbf{Receiver processing:} The signal at the input of the receiver is
\begin{equation}
    y(t)=
    \sqrt{P}\left(\sum_{k=1}^K s_k\right)\varphi(t)+w(t)
    \label{eq:rx_da}
\end{equation}

At the output of the matched filter to $\varphi(t)$, sampled at $t=T$, the signal is
\begin{equation}
    r=\sqrt{P}\sum_{k=1}^K s_k+\nu,
\end{equation}
where $\nu\sim \mathcal{N}(0,N_0/2)$. This signal is a sufficient statistic of the sample mean, which can be estimated as 
\begin{equation}
    \hat{f}=\frac{r}{\sqrt{P}K}\sim\mathcal{N}\left(
    f,\frac{N_0}{2PK^2}
    \right),
    \label{eq:linear_oac_estimate}
\end{equation}
and corresponds to the minimum variance unbiased estimator. The corresponding MSE is
\begin{equation}
    \text{MSE}_\text{c}=\frac{N_0}{2PK^2}=\frac{\alpha A^2}{2}\frac{1}{K^2}\frac{N_0}{E_s},
    \label{eq:mse_da}
\end{equation}
which depends on the source characteristics.

Note that $\text{MSE}_\text{q}=0$ because we considered an analog modulation. If the scheme is implemented using $M$-PAM, the quantization error is given by \eqref{eq:mse_q}. In this case $\text{MSE}_\text{c}$ remains the same, since the receiver performs the same estimation and does not require a digital detection step.

\section{Type-based multiple access}
\label{sec:tbma}

TBMA follows the same index modulation format described in Section \ref{sec:ortho}, where each possible data value is associated with one of the $N$ orthogonal waveforms $\varphi_n^{}$. The only difference is that TBMA does not require independent transmissions. Thus, $x_k(t)$, $E_s=P$ and $B=N/T$ remain unchanged, and the total duration time is reduced to $D=T$.\newline

\noindent
\textbf{Receiver processing}:
The received signal is
\begin{align}
    y(t)&=\sum_{k=1}^K\sum_{n=1}^N\sqrt{P}\, s_{k,n}\varphi_{n}^{}(t)+w(t)\nonumber\\
    &=\sqrt{P}\sum_{n=1}^N K_n\varphi_n^{}(t)+w(t),
\end{align}
where $K_n$ denotes the number of devices transmitting over waveform $n$. Thus, \gls{TBMA} enables over-the-air superposition when users have identical symbols.

The signal after matched filtering with $\varphi_n^{}(t)$ at sampling time $t=T$ is
\begin{equation}
    r_n=\sqrt{P}K_n +\nu_n,
    \label{eq: MF_tbma}
\end{equation}
where $\nu_n\sim\mathcal{N}(0,N_0/2)$. This signal is a sufficient statistic of the number of transmitting devices per resource,
\begin{equation}
    \tilde{K}_n=\frac{r_n}{\sqrt{P}}\sim\mathcal{N}\left(
    K_n,\frac{N_0}{2P}
    \right),
    \label{eq:users_bin_hat}
\end{equation}

The received vector $(r_1,\dots,r_{N})$ can be interpreted as a noisy unnormalized histogram (i.e., type) of the used resources. Due to the one-to-one mapping established between $\bar{d}_k$ and indices $n$, this histogram provides a noisy observation of the empirical distribution of the quantized measurements $\bar{d}_k$.

All symmetric functions can be computed directly from the histogram. In particular, the sample mean is obtained as the first moment of the histogram,
\begin{equation}
    \hat{f}=\frac{1}{K}\sum_{n=1}^N \mathcal{Q}^{-1}(n)\tilde{K}_n,
    \label{eq:TBMA_noisy}
\end{equation}
where $\mathcal{Q}^{-1}(n)$ maps the bin index to the corresponding value in the original data domain $(-A,A)$. Equation \eqref{eq:TBMA_noisy} corresponds to a nonparametric estimation and fundamentally differs from the original TBMA formulation, which is tailored to estimating a specific parameter of the data distribution. 

TBMA naturally decouples the communication problem (i.e., histogram estimation) from the computation problem (i.e., function estimation). Thus, the communication stage, which estimates the histogram, can be designed independently of the computation stage, allowing different symmetric functions to be evaluated without modifying the underlying communication protocol.

Although we assume $K$ to be known at the receiver side for simplicity, this is not a requirement and the receiver can estimate the number of transmitters as
\begin{equation}
    \hat{K}=\sum_{n=1}^N \tilde{K}_n
\end{equation}

The MSE obtained with \eqref{eq:TBMA_noisy} is
\begin{align}
    \text{MSE}_c&=\mathbb{E}\left\{\left(
    \frac{1}{K}\sum_{n=0}^{N-1}\mathcal{Q}^{-1}(n)\tilde{K}_n-\frac{1}{K}\sum_{n=0}^{N-1}\mathcal{Q}^{-1}(n){K}_n
    \right)^2\right\}\nonumber\\
    &=\frac{4A^2}{K^2N^2}\mathbb{E}\left\{\left(
    \sum_{n=0}^{N-1}n(\tilde{K}_n-K_n)
    \right)^2\right\}\nonumber\\
    &=\frac{4A^2}{K^2N^2}
    \sum_{n=0}^{N-1}n^2\frac{N_0}{2P}\nonumber\\
    &\approx \frac{2A^2}{3}\frac{N}{K^2}\frac{N_0}{E_s}
    \label{eq:mse_high_snr}
\end{align}
which scales inversely with the SNR, as in linear OAC, since channel noise affects both estimates in an additive manner. Furthermore, the additional factor of $N$ indicates that directly estimating the function from the noisy histogram yields a higher MSE than linear OAC.

A more robust estimate can be obtained by first denoising the received histogram, exploiting the fact that the histogram counts $K_n$are nonnegative integers. In particular, the contribution of user $k$ at bin $n$ is
\begin{align}
    \text{count}(k,n)&=\frac{1}{\sqrt{P}}\int_{-\infty}^\infty
    x_k(t-mT)\varphi_n^{}(t-mT)dt\nonumber\\
    &=\frac{1}{\sqrt{P}}
    \int_{-\infty}^\infty
    \sqrt{P}\varphi_{\bar{d}_k}(t-mT)\varphi_n^{}(t-mT)dt\nonumber\\
    &=\delta_{n\bar{d}_k},
    \label{eq:contribution}
\end{align}
this is, each user contributes an amplitude of $+1$ to the corresponding histogram bin.

\begin{remark} Equation \eqref{eq:contribution} depends on the adopted signal model. For example, a different normalization with respect to $P$ would modify the contributions in the histogram. Throughout the rest of the paper, we assume that each user contributes ideally with $+1$ to the corresponding histogram bin, which can be interpreted as the frequency-domain representation of the transmitted signal space.
\end{remark}

\begin{figure*}[t]
    \centering
    \includegraphics[width=0.95\textwidth]{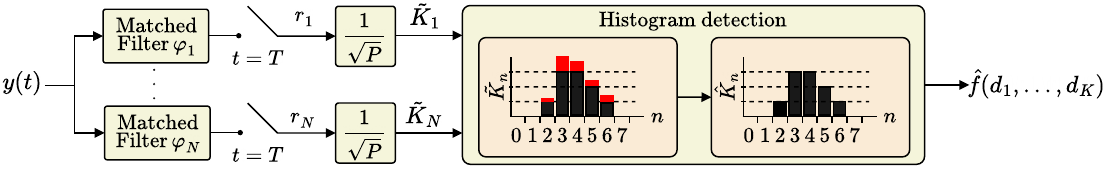}
    \vspace{5pt}
    \caption{Receiver architecture for separate histogram detection and function computation with TBMA. The clean histogram $(\hat{K}_1,\dots,\hat{K}_N)$ is recovered from the noisy observations $(\tilde{K}_1,\dots,\tilde{K}_N)$ using one of the two proposed detectors.}
  \label{fig:lattice_decoder}
\end{figure*}

Figure \ref{fig:lattice_decoder} illustrates the receiver architecture, which first obtains the noisy histogram $(\tilde{K}_1,\dots,\tilde{K}_N)$, and then applies a digital detector to recover the denoised histogram $(\hat{K}_1,\dots,\hat{K}_N)$. In the following, we present and compare two detection strategies.

\subsection{Independent histogram detection}
Since $K_n\in \mathbb{Z}_+$, a denoised histogram may be obtained by projecting each $\tilde{K}_n$ onto the nearest integer. From a communication perspective, this is equivalent to detecting a noisy PAM symbol with constellation $\mathbb{Z}_+$. Therefore, the ML detector reduces to minimum-distance decoding,
\begin{equation}
    \hat{K}_n=\argmin_{k\in\mathbb{Z}_+} \left(\tilde{K}_n-k\right)^2=\bigl \lfloor
    \tilde{K}_n
    \bigr \rfloor,
\end{equation}
which simplifies to rounding and clipping to zero if necessary. After rounding all $N$ bins individually, equation \eqref{eq:TBMA_noisy} is used to compute the sample mean.

Since the behavior of this scheme strongly depends on SNR, we will compute the MSE in two separate regimes: high SNR, where rounding effectively eliminates most small errors, and low SNR, where noise dominates and rounding has little effect.\newline

\noindent
\textbf{High SNR:} At sufficiently large SNR, the noise component becomes small relative to the symbol spacing, and detection errors are dominated by nearest-neighbor transitions. Therefore, the additive noise only pushes the observation to an adjacent integer values, causing errors of the form $\hat{K}_n=K_n\pm1$. In these cases, the SER at resource $n$ corresponds to
\begin{align}
    P_{e,n} &\approx 1-
    \text{Pr}\left(K_n-\frac{1}{2}\leq\hat{K}_n\leq K_n+\frac{1}{2}\right)\nonumber\\
    &=1-
    \text{Pr}\left(-\frac{1}{2}\leq \frac{\nu_n}{\sqrt{P}}\leq \frac{1}{2}\right)\approx2Q\left(\sqrt{\frac{E_s}{2N_0}}\right),
\end{align}
which corresponds to the SER of $M$-PAM with distance of 1 between adjacent symbols. 

Each detection error results in a squared error of one, while correct detections yield no error. Thus, the MSE at resource $n$ is
\begin{align}
    \text{MSE}_{\text{c},n}=\mathbb{E}\left\{(\hat{K}_n-K_n)^2\right\}=
    (K_n\pm1-K_n)^2P_{e,n}=P_{e,n},
\end{align}
and the overall channel-induced distortion becomes
\begin{align}
    \text{MSE}_\text{c}&=\mathbb{E}\left\{\left(
    \frac{1}{K}\sum_{n=1}^{N}\mathcal{Q}^{-1}(n)\hat{K}_n-\frac{1}{K}\sum_{n=1}^{N}\mathcal{Q}^{-1}(n){K}_n
    \right)^2\right\}\nonumber\\
    &=\mathbb{E}\left\{\left(
    \frac{1}{K}\sum_{n=1}^{N}\mathcal{Q}^{-1}(n)(\hat{K}_n-K_n)
    \right)^2\right\}\nonumber\\
    &=\frac{1}{K^2}
    \sum_{n=1}^{N}\left(\mathcal{Q}^{-1}(n)\right)^2\text{MSE}_{\text{c},n}\nonumber\\
    &= \frac{2A^2(N^2-1)}{3K^2N}
    Q\left(\sqrt{\frac{E_s}{2N_0}}\right)\nonumber\\
    &\approx \frac{2A^2}{3}\frac{N}{K^2}
    Q\left(\sqrt{\frac{E_s}{2N_0}}\right),
    \label{eq:mse_c_integer}
\end{align}
which decays exponentially with the SNR due to the exponential decay of the $Q$-function.\newline


\noindent
\textbf{Low SNR:} Noise dominates the rounding operation, and the estimation error can therefore be approximated as
\begin{equation}
    \hat{K}_n-K_n =
    \Biggl \lfloor
    K_n +\frac{\nu_n}{\sqrt{P}}
    \Biggr \rfloor - K_n=
    \Biggl \lfloor
    \frac{\nu_n}{\sqrt{P}}
    \Biggr \rfloor
    \approx \frac{\nu_n}{\sqrt{P}},
\end{equation}
where the noise introduced by rounding is negligible. The resulting MSE is given by \eqref{eq:mse_high_snr}, indicating that TBMA requires a minimum SNR to achieve reliable estimation and outperform linear OAC.

\subsection{ML Histogram Detection}
The previous detector is suboptimal because each histogram component is decoded independently and the global constraints defining a valid histogram are not exploited. In this section, we exploit this joint structure and formulate histogram recovery as a constrained ML detection problem.

From a communication perspective, recovering the transmitted histogram corresponds to jointly detecting the histogram $(K_1,\ldots,K_N)$ from the noisy observation $(\tilde{K}_1,\ldots,\tilde{K}_N)$. According to \eqref{eq:users_bin_hat}, each observation is corrupted by AWGN, and the ML detector reduces to the minimum Euclidean distance rule:
\begin{subequations}
\begin{align}
\nonumber
& \underset{k_1,\dots,k_N}{\text{minimize}} ~~~~~~
\,\sum_{n=1}^N J_n(k_n)=\sum_{n=1}^N (\tilde{K}_n-k_n)^2 \\
& \hspace{-20pt}\qquad\text{subject to}~~~~~~
\sum_{n=1}^N k_n=K\label{const:sum_k_1}\\
&~~~~~~~~~~~~~~~~~~
k_n\in\mathbb{Z}_+,\forall n\label{const:nonneg}
\end{align}
\label{eq:min_mse}
\end{subequations}
where $J_n(k_n)$ denotes the contribution of resource $n$ with occupancy $k_n$, constraint \eqref{const:sum_k_1} enforces the allocation of exactly $K$ users across $N$ resources, and constraint \eqref{const:nonneg} enforces nonnegative integer occupancies.

For given $K$ and $N$, the cardinality of the feasible search space is
\begin{equation}
\binom{K+N-1}{N-1}=\frac{(K+N-1)!}{K!(N-1)!},
\end{equation}
which grows dramatically fast and makes exhaustive search computationally prohibitive. Nevertheless, in the following, we prove that the ML detector  admits an exact greedy implementation. Theorem \ref{thm:greedy_optimality} formalizes the optimal procedure.

\begin{theorem}[Optimality of Greedy ML Detection]
\label{thm:greedy_optimality}
Initialize $k_n=0, \forall n$. At each iteration, select the resource with the largest occupancy,
\begin{equation}
    n^\star
    =
    \arg\max_{n}
    \tilde{K}_n,
    \label{eq:greedy_selection}
\end{equation}
assign one additional user to resource $n^\star$,
\begin{equation}
    k_{n^\star}\leftarrow k_{n^\star}+1,
\end{equation}
and remove it from the histogram,
\begin{equation}
    \tilde{K}_{n^\star}\leftarrow \tilde{K}_{n^\star}-1
\end{equation}
After $K$ iterations, the resulting allocation
$(k_1,\dots,k_N)$ is globally optimal.
\end{theorem}

\begin{proof}
The proof is presented in the Appendix.
\end{proof}

\begin{figure*}[t]
    \centering
    \includegraphics[width=\textwidth]{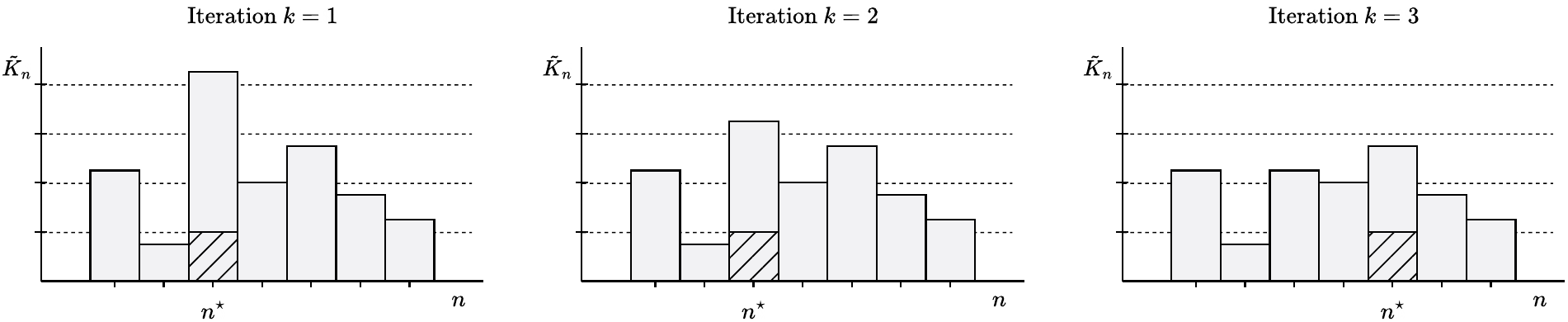}
  \caption{Example of the proposed successive cancellation decoder over three iterations. At each iteration, the decoder identifies the histogram bin with the largest value ($n^\star$), decodes one user assigned to that bin (hatch), and subtracts it from the corresponding bin. After three iterations, the recovered resource indices are $\bar{d}_k=\{3,3,5\}$.}
  \label{fig:SIC_decoder}
\end{figure*}

This procedure admits a natural interpretation as successive cancellation, in which one signal component is detected and removed before proceeding to the next detection. Unlike conventional successive interference cancellation (SIC) receivers employed in wireless communications \cite{verdu1998multiuser}, the proposed approach is ML-optimal. The reason is that each detected signal has a known and deterministic contribution according to \eqref{eq:contribution}, and no amplitude estimation is required. Moreover, the receiver does not attempt to identify the device associated with each detection, which is not a requirement for OAC, and any arbitrary order is sufficient for accurate function computation. Figure \ref{fig:SIC_decoder} exemplifies the procedure.

The greedy selection also admits an intuitive signal processing interpretation. According to \eqref{eq: MF_tbma}, the SNR per resource is proportional to $K_n$. Therefore, decoding the histogram components in descending order of magnitude prioritizes those with higher SNR and, consequently, lower SER. Additionally, this strategy does not require the receiver to know the total number of devices $K$ in advance. Alternatively, the receiver can terminate the decoding process when the power difference between the maximum and minimum bins falls below a predefined threshold.\newline

\noindent
\textbf{SER:} To determine the theoretical performance of the ML decoder, we use a conventional digital communication approach. The histogram is an $N$-dimensional symbol with an associated SER. We consider a high-SNR regime, so that decoding errors are dominated by nearest-neighbor events, and the histogram SER can be upper bounded with the union bound,
\begin{equation}
    \text{Pr}
    \left(
    \exists!\, n \,|\, \hat{K}_n\neq K_n
    \right)\leq
    N_\text{nn}Q
    \left(
    \sqrt{\frac{d_\text{min}^2}{4\sigma_w^2}}
    \right),
    \label{eq:SER_histogram}
\end{equation}
where $N_\text{nn}$ denotes the number of nearest-neighbor histograms at Euclidean distance $d_\text{min}$ and $\sigma_w^2$ is the noise variance of $\tilde{K}_n$ in \eqref{eq:users_bin_hat}.

For histograms, a nearest-neighbor corresponds to moving a single contribution from the correct bin to another while leaving all remaining contributions unchanged. Without loss of generality, consider a user moved from the first to the second bin. The corresponding minimum Euclidean distance is
\begin{equation}
    d_\text{min}=||(-1,1,0,\dots,0)||=\sqrt{2}
    \label{d_min}
\end{equation}

The number of nearest neighbors is obtained by selecting one occupied bin as the source of the user transfer and one of the remaining $N-1$ bins as the destination. We upper-bound the number of occupied bins with $\min(K,N)$: When $K<N$, the maximum is attained when every user occupies a different bin, whereas for $K\geq N$ all bins can be occupied. Thus, we can upper-bound the number of nearest neighbors by
\begin{equation}
    N_\text{nn}\leq (N-1)\min(K,N),
    \label{eq:neighbors}
\end{equation}

Substituting \eqref{d_min} and \eqref{eq:neighbors} into \eqref{eq:SER_histogram} gives
\begin{equation}
    \text{Pr}
    \left(
    \exists!\, n \,|\, \hat{K}_n\neq K_n
    \right)\leq
    N\min(K,N)Q
    \left(
    \sqrt{\frac{E_s}{N_0}}
    \right)
    \label{eq:SER_histogram2}
\end{equation}
\newline

\noindent
\textbf{MSE:} Conditioned on a decoding error with a nearest-neighbor, the MSE can be approximated as
\begin{align}
    &\hspace{-30pt}\mathbb{E}\bigg\{
    \left(\bar{f}-\hat{f}\right)^2
    \Big|
    \hat{K}_n\neq K_n,\exists! n
    \bigg\}\nonumber\\
    &\quad\approx
    \mathbb{E}\bigg\{
    \left(\bar{f}-\hat{f}\right)^2
    \Big|
    \bar{d}_1\neq\mathcal{Q}\left(\hat{d}_1\right)
    \bigg\}\nonumber\\
    &\quad=\frac{4A^2}{N^2K^2}\mathbb{E}\left\{\left(
    \hat{d}_1-\bar{d}_1
    \right)^2\right\}\nonumber\\
    &\quad=\frac{2A^2}{3K^2},
    \label{eq:mse_bound2}
\end{align}
which follows from \eqref{eq:mse_bound_OMA}. Finally, the channel-induced MSE is
\begin{align}
    \text{MSE}_\text{c}&=
    \mathbb{E}\bigg\{
    \left(\bar{f}-\hat{f}\right)^2\Big|
    \hat{K}_n\neq K_n, \exists!\, n
    \bigg\}
    \text{Pr}
    \left(
    \exists!\, n \,|\, \hat{K}_n\neq K_n
    \right)\nonumber\\
    &\approx
    \frac{2A^2}{3}
    \frac{N\min(K,N)}{K^2}
    Q
    \left(
    \sqrt{\frac{E_s}{N_0}}
    \right)
    \label{eq:mse_conditioned}
\end{align}

Notice that this bound scales less favorably than the independent histogram detection in \eqref{eq:mse_c_integer} because of the increased number of nearest-neighbors. However, this penalty is offset by the double SNR argument of the $Q$-function, which arises from exploiting the orthogonality of the signaling waveforms.


\subsection{Discussion}
\label{sec:discussion}

\begin{table*}[t]
\centering
\small
\setlength{\tabcolsep}{12pt}
\renewcommand{\arraystretch}{1.25}
\caption{Comparison of different OAC schemes.}
\begin{tabular}{c| c c c c c}
\toprule
\textbf{Scheme (Modulation)} &
\textbf{Duration} &
\textbf{Bandwidth} &
$E_s$ &
$\text{MSE}_\text{q}$ &
$\text{MSE}_\text{c}$ \\
\midrule

\textbf{TDMA (M-FSK)}
& $KT$
& $\dfrac{N}{T}$
& $P$
& $\dfrac{A^2}{3N^2K}$
& $\dfrac{2A^2}{3}\dfrac{N}{K}Q\left(\sqrt{\dfrac{E_s}{N_0}}\right)$
\\
\midrule

\textbf{Linear OAC (DSB)}
& $T$
& $\dfrac{1}{T}$
& $\alpha A^2P$
& $0$
& $\dfrac{\alpha A^2}{2}\dfrac{1}{K^2}\dfrac{N_0}{E_s}$
\\
\midrule

\makecell{\textbf{TBMA (M-FSK)}\\
Independent detection}
& $T$
& $\dfrac{N}{T}$
& $P$
& $\dfrac{A^2}{3N^2K}$
& $\dfrac{2A^2}{3}\dfrac{N}{K^2}
Q\left(\sqrt{\dfrac{E_s}{2N_0}}\right)$
\\
\midrule

\makecell{\textbf{TBMA (M-FSK)}\\
Joint detection}
& $T$
& $\dfrac{N}{T}$
& $P$
& $\dfrac{A^2}{3N^2K}$
& $\dfrac{2A^2}{3}\dfrac{N\min(K,N)}{K^2}
Q\left(\sqrt{\dfrac{E_s}{N_0}}\right)$
\\
\bottomrule
\end{tabular}
\label{tab:comparison_schemes}
\end{table*}

Table \ref{tab:comparison_schemes} summarizes the communication schemes studied in this work: 
\begin{itemize}
    \item \textbf{TDMA:} orthogonal multiple access from section \ref{sec:ortho}. For each orthogonal transmission, the point-to-point communication is implemented with an M-FSK index modulation.
    \item \textbf{Linear OAC:} conventional analog modulation from section \ref{sec:ba}, and implemented with DSB.
    \item \textbf{TBMA:} proposed type-based OAC from section \ref{sec:tbma}, and implemented with M-FSK.
\end{itemize}

The schemes are compared in terms of total transmission duration, total bandwidth, per-user energy consumption and channel-induced MSE. TDMA exhibits the least favorable time–bandwidth trade-off, whereas TBMA requires $N$ times more bandwidth than linear OAC. In terms of energy consumption, TBMA has a fixed per-user energy cost, while linear OAC depends on the transmitted data. These trade-offs also apply when implementing the schemes using time rather than frequency resources. For example, with pulse-position modulation (PPM), reducing the bandwidth comes at the expense of a proportionally longer transmission duration, and vice versa. However, this exchange between time and bandwidth does not affect the channel-induced MSE, provided that the resources remain orthogonal.

The analytical MSE expressions reveal three key factors that determine performance:
\begin{itemize}
    \item The first term captures the impact of the data distribution. While all methods depend on the data range as $A^2$, linear OAC depends on the variance parameter $\alpha$.
    
    \item The second term characterizes the trade-off between the number of users, $K$, and the number of communication resources, $N$. Linear OAC is independent of $N$, as it requires one single resource. In contrast, TBMA with joint detection scales as $N/K$ when $N>K$ and as $N^2/K^2$ when $N<K$. Since TDMA scales as $N/K$, TBMA is always equal or better than TDMA, which highlight the advantage in terms of resource efficiency as the number of users increases.
    
    \item The third term describes the dependence of the MSE on the SNR. Linear OAC exhibits the least favorable behavior, while TBMA is expected to benefit more significantly from increasing SNR.
\end{itemize}

An additional advantage of TBMA is its compatibility with existing communication technologies, such as Long Range communications (LoRa) and orthogonal frequency division multiplexing (OFDM), which already support orthogonal signaling. This enables TBMA to leverage established transceiver architectures, avoiding the modifications often required by linear OAC schemes.

The analysis presented so far focuses on linear function computation, for which the MSE expressions can be characterized analytically. Extending this analysis to nonlinear functions requires numerical evaluations, since the corresponding MSE expressions depend on the specific function being evaluated and may not admit closed-form solutions.




\section{Numerical Results}
\label{sec:results}

In this section, we numerically evaluate the performance of the different OAC schemes presented in Table \ref{tab:comparison_schemes} for both linear and nonlinear functions. We first consider $d_k\sim\mathcal{U}(-1,1)$, which implies in $A=1$ and $\alpha=1/3$. 

\begin{figure}[t]
    \centering
    \includegraphics[width=\columnwidth]{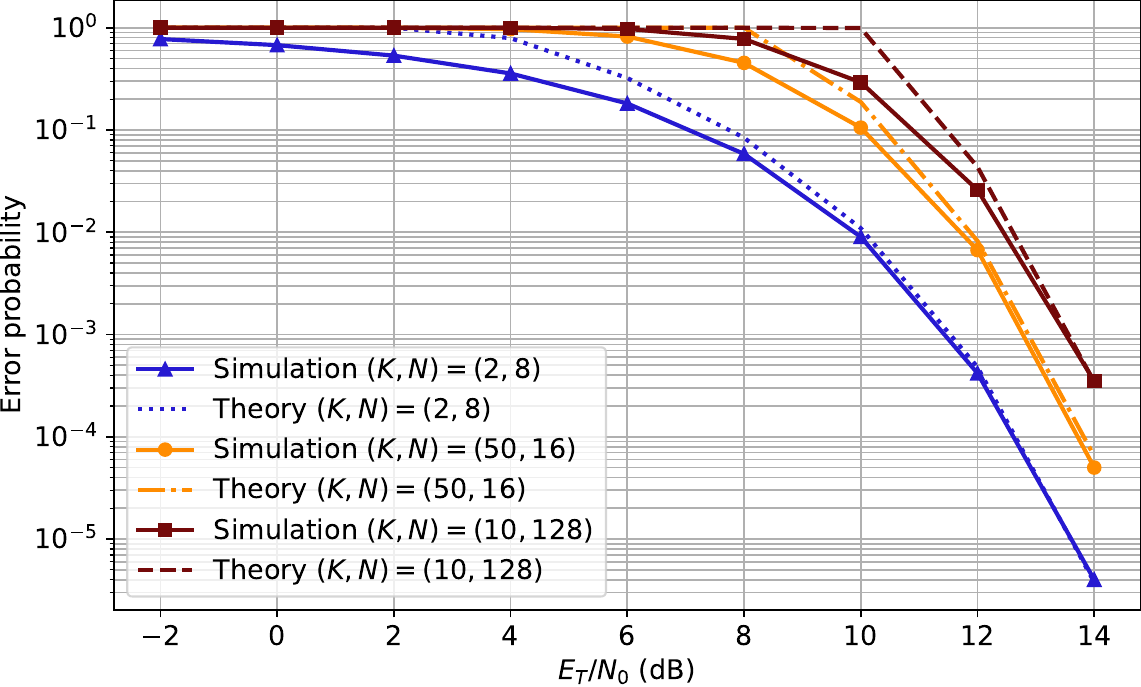}
    \caption{Theoretical and empirical SER of joint ML histogram detection under different $(K,N)$ configurations.}
    \label{fig:error_prob}
\end{figure}

Figure \ref{fig:error_prob} compares the theoretical and empirical SER of joint ML histogram detection for different values of $(K,N)$. Since deriving an exact expression for the SER is intractable, the theoretical curves are obtained using the upper bound in \eqref{eq:SER_histogram2}. Detection is performed using the proposed successive cancellation algorithm. The considered configurations yield $N_\text{nn}=16$, $256$ and $1280$, respectively, illustrating the impact of $(K,N)$ on the SER. The close agreement between the simulation and the theoretical bound at $E_s/N_0\geq 10$ dB supports the optimality of the greedy decoder.

\begin{figure}[t]
    \centering
    \includegraphics[width=\columnwidth]{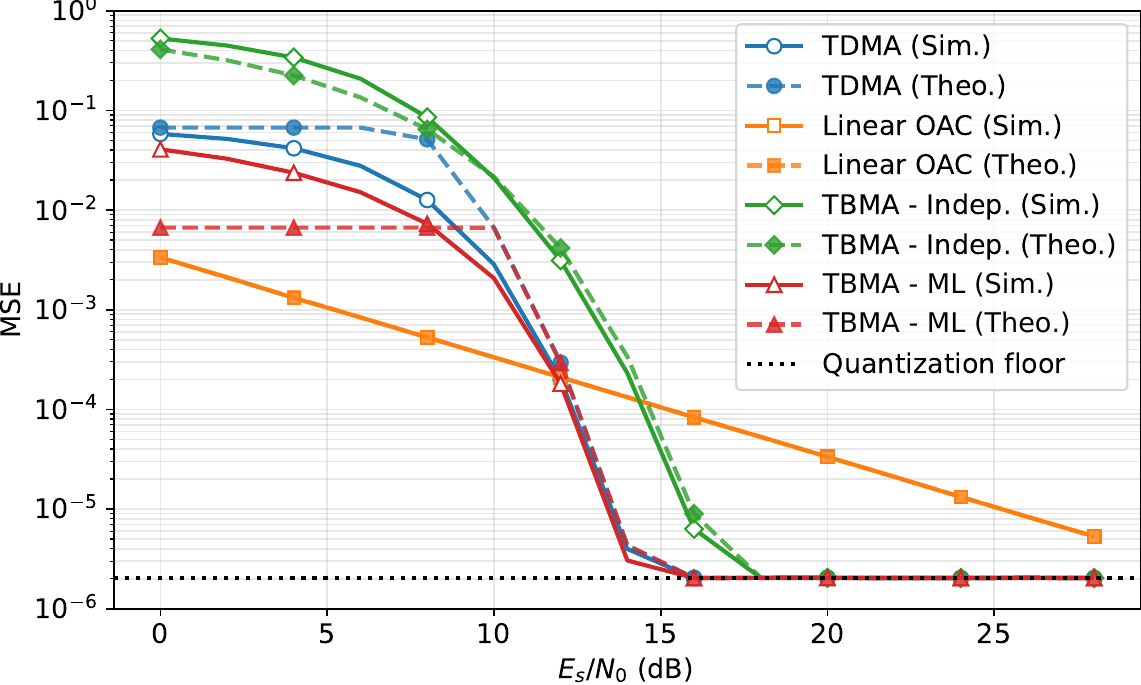}
    \caption{Theoretical and empirical MSE of the sample mean function for $x_k\in\mathcal{U}(-1,1)$ and $(K,N)=(10,128)$.}
    \label{fig:mse_mean}
\end{figure}

Figure \ref{fig:mse_mean} compares the theoretical and simulated MSE of the sample mean function for $(K,N)=(10,128)$ across all schemes. The choice of a large $N$ allows digital OAC to achieve a low quantization noise, while setting $K<N$ illustrates that TBMA can operate effectively without requiring a large number of superpositions. For TBMA, the theoretical bounds closely match the simulated results for $E_s/N_0\geq 10$ dB, consistent with the high-SNR regime identified in Figure \ref{fig:error_prob}. Successive cancellation achieves a lower MSE than independent bin detection, with only a modest increase in computational complexity. The gain is asymptotically upper-bounded by 3 dB, corresponding to the factor o 2 difference in the SNR arguments.
Notably, successive cancellation achieves the same performance as TDMA while allowing all $K$ users to transmit simultaneously. Therefore, as illustrated in Table \ref{tab:comparison_schemes}, TBMA occupies the same bandwidth and reduces the total transmission duration by a factor of $K$ without degrading estimation performance. Finally, linear OAC exhibits a linear improvement, without an error floor, while using a single communication resource. These results establish linear OAC as the preferred approach for linear function computation.

\begin{figure}[t]
    \centering
    \includegraphics[width=\columnwidth]{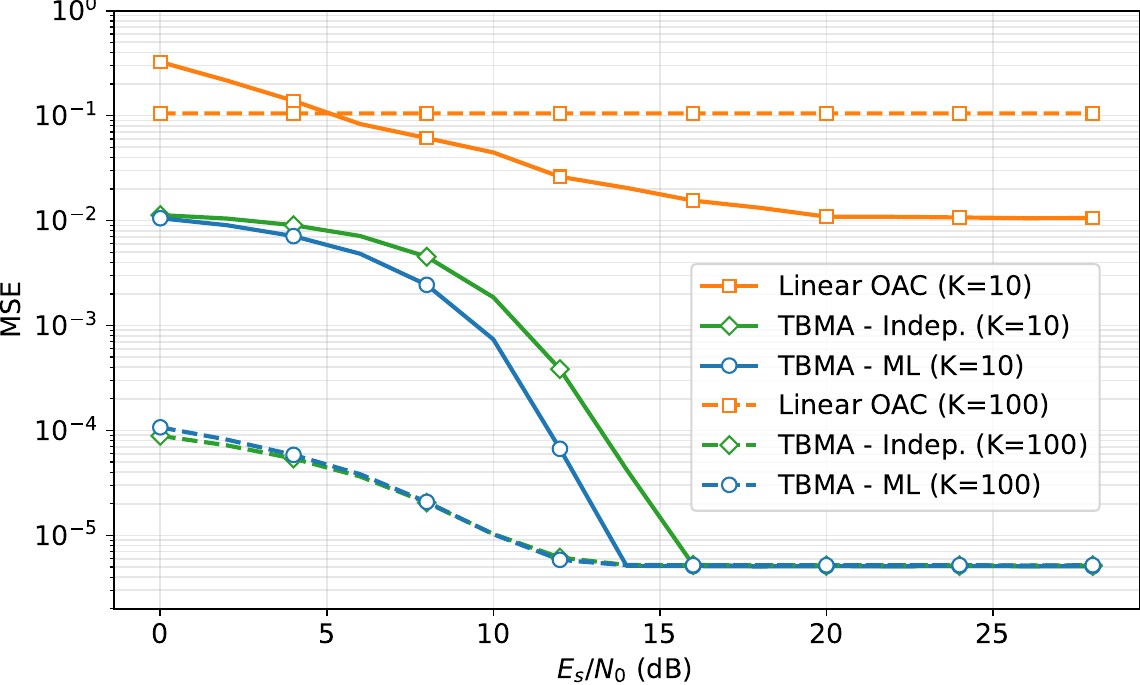}
    \caption{Empirical MSE of the maximum function for \mbox{$x_k\in\mathcal{U}(0,1)$}, $N=128$ and $K=\{10,100\}$.}
    \label{fig:mse_max}
\end{figure}

We next evaluate nonlinear function computation. We consider inputs $x_k\sim\mathcal{U}(0,1)$, otherwise there are no feasible nomographic decompositions and linear OAC cannot be realized. Figure \ref{fig:mse_max} compares the empirical MSE of linear OAC and TBMA for maximum estimation, and also illustrate the impact of $K$. In linear OAC, the maximum is approximated using the soft-max function,
\begin{equation}
    \max\{x_1,\dots,x_K\}\approx\frac{1}{\beta}\ln\left(\sum_{k=1}^{K}e^{\beta x_k}\right),
\end{equation}
which is parametrized by $\beta$. We set $\beta=8$ in our simulations. It can be shown that the soft-max approximation produces a bias bounded by $\ln(K)/\beta$. Thus, increasing $\beta$ improves the approximation but also increases the average transmit power, creating a trade-off between estimation accuracy and power consumption. Moreover, the logarithmic post-processing alters the AWGN assumption and makes the estimate more sensitive to noise. Together, these effects contribute to the relatively poor performance of linear OAC observed in Figure \eqref{fig:mse_max}. More importantly, the $1/K^2$ MSE scaling observed for linear functions does not carry over to general function computation because of the approximation bias. In contrast, TBMA exhibits the same exponential decay observed in linear functions because it decouples the communication and computation stages. Similarly, the exponential decay is observed around 12 dB. Their performance also improve with increasing $K$, consistent with the linear function computation.

\begin{figure}[t]
    \centering
    \includegraphics[width=\columnwidth]{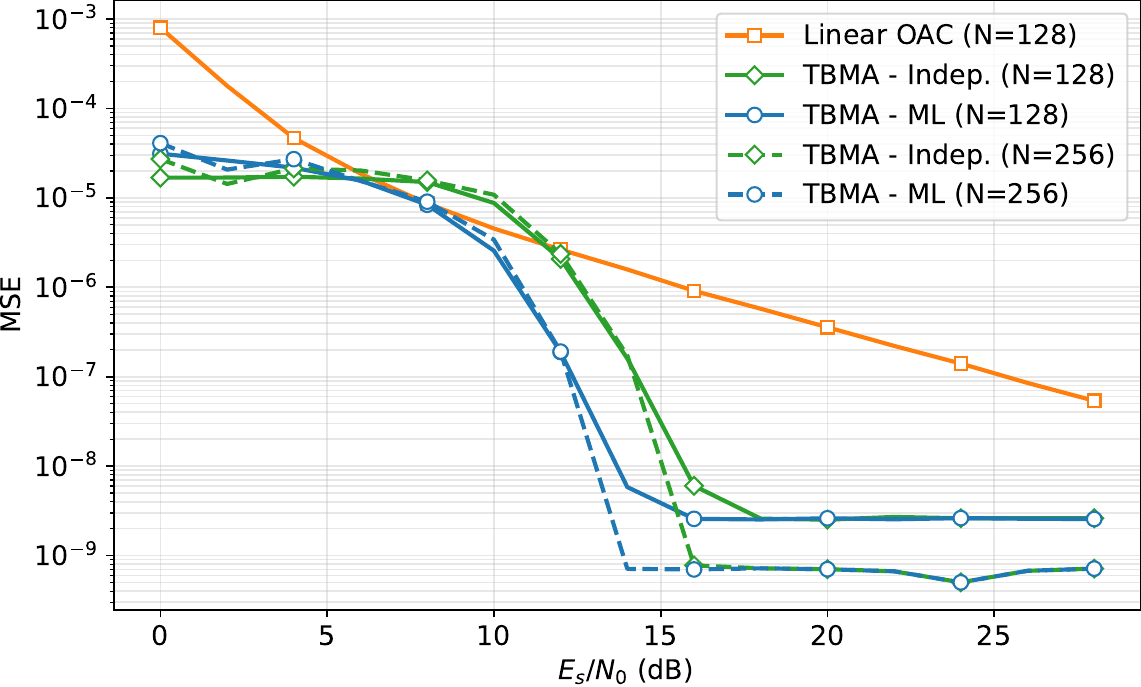}
    \caption{Empirical MSE of the product function for \mbox{$x_k\in\mathcal{U}(0,1)$}, $N=\{128,512\}$ and $K=10$.}
    \label{fig:mse_prod}
\end{figure}

Figure \ref{fig:mse_prod} illustrates the empirical MSE for computing the product with $x_k\sim\mathcal{U}(0,1)$, and the impact of $N$. Linear OAC uses the nomographic decomposition in \eqref{eq:prod}, whereas TBMA estimates it directly from the detected histogram. These results illustrate the trade-off between resource consumption and computation accuracy seen in Table \ref{tab:comparison_schemes} for TBMA. Increasing $N$ reduces the quantization-induced MSE, at the cost of requiring greater bandwidth. This trade-off is analogous to that in conventional communication systems, where additional bandwidth can be exchanged for reduced transmit power.

\begin{figure}[t]
    \centering
    \includegraphics[width=0.95\columnwidth]{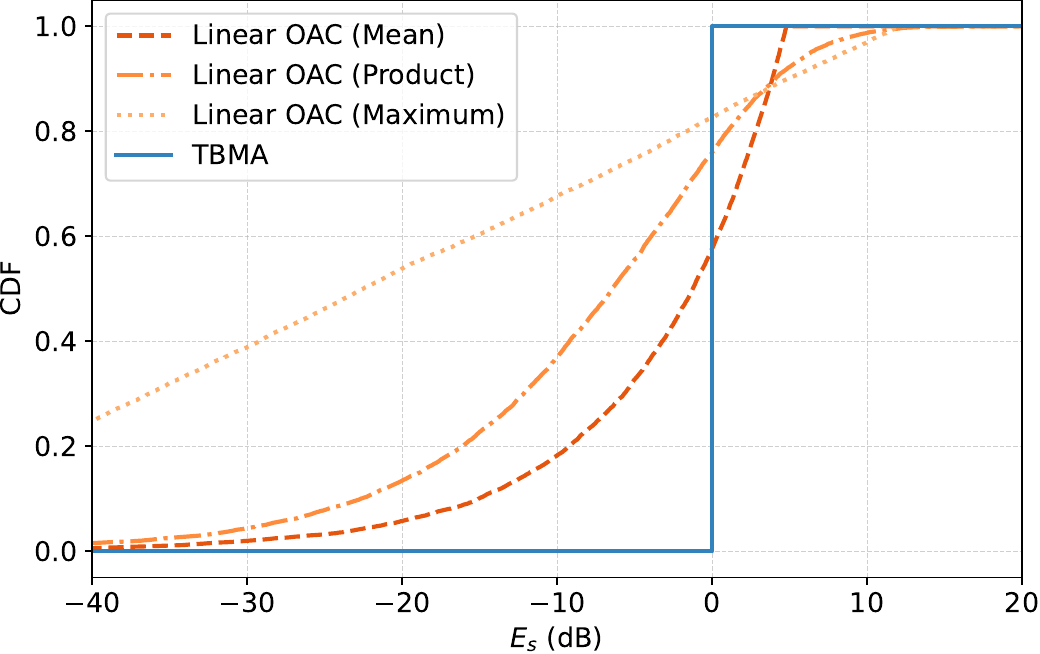}
    \caption{Cumulative distribution function of the energy consumption per-user for the sample mean, product and maximum functions.}
    \label{fig:energy}
\end{figure}

Figure \ref{fig:energy} illustrates the distribution of per-user transmit energy $E_s$ for linear OAC and TBMA when computing the previously considered functions. The transmit power $P$ is set such that the average energy per user is $\mathbb{E}\{E_s\}=1$. The results show that the energy distribution of linear OAC depends strongly on the preprocessing function. Specifically, approximately 60\%, 75\%, and 83\% of users transmit above the average energy when computing the mean, product and maximum functions, respectively. This behavior reflects the unequal energy allocation induced by the corresponding pre-processing functions, and a similar effect would be observed when the underlying data distribution changes. In contrast, TBMA uses constant-energy transmissions, no matter the function being computed. This uniform energy allocation can be particularly advantageous when devices may have limited and heterogeneous energy resources.

\begin{figure}[t]
    \centering
    \includegraphics[width=\columnwidth]{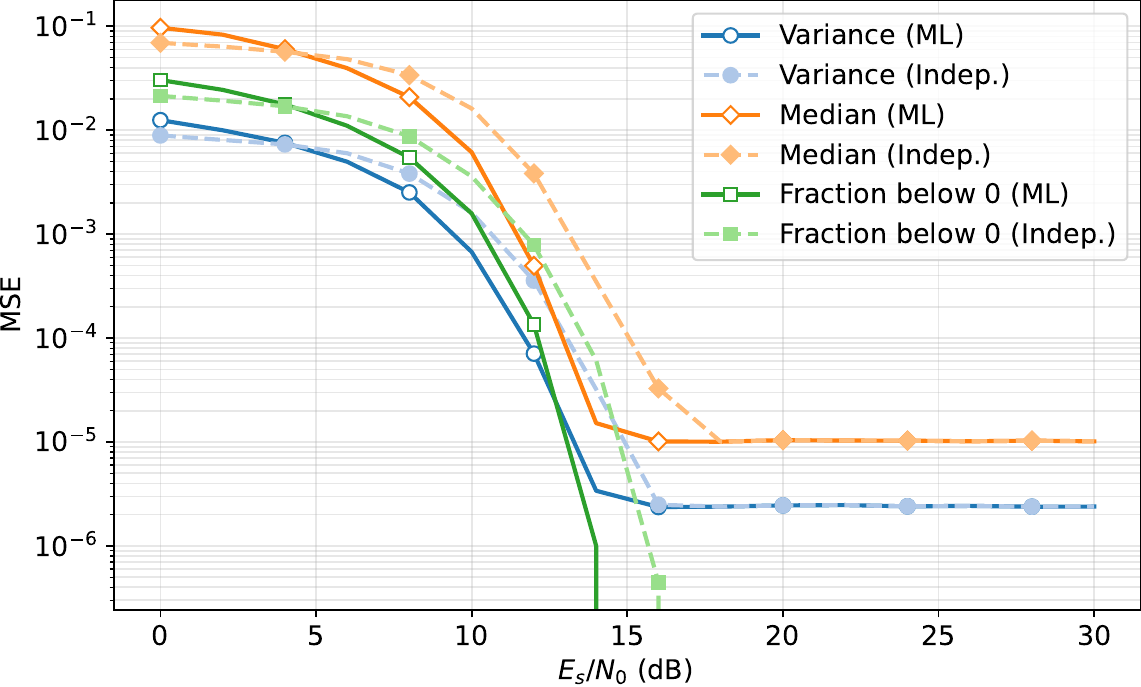}
    \caption{Empirical MSE of TBMA for alternative nonlinear functions with \mbox{$x_k\in\mathcal{U}(-1,1)$}, $N=128$ and $K=10$.}
    \label{fig:other_functions}
\end{figure}

Finally, we show the potential of having the histogram as an intermediate representation step. Specifically, Figure \ref{fig:other_functions} illustrates the performance of TBMA for computing three functions: variance, median and the fraction of users below a threshold. These functions are less common in the OAC literature because they cannot be directly computed with linear OAC, since no suitable nomographic decompositions are known. As expected, the MSE decreases with increasing SNR for all three functions. Notably, the fraction of users below a threshold exhibits no quantization error floor, since its output is an integer number. More importantly, all three functions can be computed from a single transmission, a capability not readily supported by existing OAC frameworks. This is feasible because TBMA keeps the transmission and receiver processing independent of the target function, enabling multiple functions to be computed from the same received histogram.

\section{Conclusion}
This paper has investigated TBMA as a general framework for OAC, in which the coherent superposition of orthogonal modulations enables the receiver to recover a histogram of distributed data. TBMA decouples the communication scheme from the function to be computed, allowing a common transmission architecture to support a broad class of functions without requiring function-specific designs. We have shown that TBMA can be implemented using M-FSK and PPM, and is supported by practical communication technologies such as LoRa and OFDM.

We formulated type-based OAC as a histogram detection problem and proposed two detectors: independent bin quantization and joint ML detection. We showed that optimal ML detection admits a low-complexity greedy algorithm based on successive cancellation. 
We established a systematic comparison between TBMA, orthogonal multiple access and linear OAC in terms of MSE, transmission time, bandwidth and energy consumption. The numerical results further validate the theoretical findings and illustrate the superiority of TBMA as a tool for general nonlinear function computation. Across all experiments, TBMA exhibits an exponential improvement of the MSE with the SNR and constant energy consumption.

Although TBMA utilizes more communication resources, the histogram represents a flexible tool that enables complex decision-making tasks beyond the capabilities of existing physical-layer techniques. It is particularly relevant when the target function changes over time or when several aggregate quantities must be computed from the same observations. For instance, TBMA has the potential to enhance resilience against Byzantine attacks at the physical layer, as anomalous contributions are easy to identify and mitigate. Overall, OAC is well positioned to emerge as a specialized capability that complements conventional multiple access schemes, delivering unique benefits in applications where in-network processing is particularly advantageous.

\appendix

\label{app:proof_thm}

The next two lemmas characterize the structure of the objective function and provide the basis for the theorem.

\begin{lemma}[Marginal Gain Structure]
\label{lem:marginal_gain}
Let $G_{n,k}$ denote the reduction in the objective function $J_n$ obtained from increasing the occupancy from $k-1$ to $k$ users:
\begin{equation}
    G_{n,k}=J_n(k_n-1)-J_n(k_n)=2(\tilde{K}_n-k)+1
    \label{eq:gain}
\end{equation}
The marginal gains are strictly decreasing in $k$:
\begin{equation}
    G_{n,1}>G_{n,2}>\dots>G_{n,K}
    \label{eq:monotone_gains}
\end{equation}

\begin{proof}
From the definition of $J_n(k)$, equation \eqref{eq:gain} is straightforward. Then, the difference between two consecutive gains is
\begin{equation}
G_{n,k}-G_{n,k+1}=2>0,
\end{equation}
which proves the strict monotonicity.
\end{proof}
\end{lemma}

\begin{lemma}[Marginal Gain Decomposition]
\label{lem:gain_decomposition}
The objective function in \eqref{eq:min_mse} can be expressed as
\begin{equation}
\sum_{n=1}^{N}J_n(k_n)
=
\sum_{n=1}^{N}J_n(0)
-
\sum_{n=1}^{N}\sum_{i=1}^{k_n}G_{n,i},
\label{eq:cost_gains}
\end{equation}
that is, the initial cost corresponding to zero allocated users minus the accumulated marginal gains.

\begin{proof}
The cost associated with
resource $n$ can be recursively expressed as
    \begin{align}
    J_n(k_n)&=J_n(k_n-1) - G_{n,k}\nonumber\\
    &= J_n(k_n-2) - G_{n,k-1} - G_{n,k}\nonumber\\
    &= J_n(0) - \sum_{i=1}^{k_n} G_{n,i},
    \label{eq:gain_recursive}
\end{align}
and summing over all resources yields equation \eqref{eq:gain_recursive}.
\end{proof}
\end{lemma}

Next, define $z_{n,k}\in\{0,1\}$, where $z_{n,k}=1$ indicates that resource $n$ is assigned $k$ users. From Lemma \ref{lem:gain_decomposition}, the cost function can be expressed as
\begin{equation}
    \sum_{n=1}^N J_n(k_n)=\sum_{n=1}^N J_n(0) -
    \sum_{n=1}^N \sum_{k=1}^K z_{n,k}G_{n,k}
    \label{eq:cost_gains_v2}
\end{equation}

Since the first term in \eqref{eq:cost_gains_v2} is constant, minimizing the Euclidean distance is equivalent to maximizing the accumulated gain. Consequently, problem \eqref{eq:min_mse} can be reformulated as
\begin{subequations}
\begin{align}
\nonumber
& \underset{\{z_{n,k}\}}{\text{maximize}} ~~~~~~
\,\sum_{n=1}^N\sum_{k=1}^K z_{n,k}G_{n,k} \\
& \hspace{-20pt}\qquad\text{subject to}~~~~~~
\sum_{n=1}^N\sum_{k=1}^K z_{n,k}=K\label{const:sum_k_2}\\
&~~~~~~~~~~~~~~~~~~
z_{n,k}\in\{0,1\},\forall n,k\label{const:bool}\\
&~~~~~~~~~~~~~~~~~~
z_{n,k+1}\leq z_{n,k},\forall n,k\label{const:decrease}
\end{align}
\label{eq:max_gain}
\end{subequations}

Constraints \eqref{const:sum_k_2} and \eqref{const:bool} replace \eqref{const:sum_k_1} and \eqref{const:nonneg}, respectively. Constraint \eqref{const:decrease} enforces the monotonicity of Lemma \ref{lem:marginal_gain}, since selecting the $k$th gain of a given bin implies that all preceding gains from the same bin have already been selected. In other words, assigning the $k$-th user to a bin necessarily implies the presence of the preceding $k-1$ users.

For exposition, collect all possible gains into the matrix
\begin{equation}
    \mathbf{G}=
     2\begin{bmatrix}
        \tilde{K}_1-1 & \tilde{K}_1-2 & \tilde{K}_1-3 & \cdots \\
        \tilde{K}_2-1 & \tilde{K}_2-2 & \tilde{K}_2-3 & \cdots \\
        \vdots & \vdots & \vdots & \\
        \tilde{K}_N-1 & \tilde{K}_N-2 & \tilde{K}_N-3 & \cdots \\
    \end{bmatrix}+1
    \label{eq:gain_matrix}
\end{equation}

By Lemma~\ref{lem:marginal_gain}, the marginal gains in each row of $\mathbf{G}$ are strictly decreasing from left to right. Therefore, at the beginning, only the first entry in each row (i.e., the first column) is eligible for selection. The largest eligible gain is obtained from the element with the largest observed occupancy, i.e., $n^\star=\arg\max_n \tilde{K}_n$, and the corresponding bin is assigned one user, $k_{n^\star}=1$.

After this assignment, the next eligible gain for this row is the second entry, due to monotonicity. This corresponds to assigning a second user to the same bin. Hence, the greedy procedure advances to the next entry in the selected row, while the candidate entry in every other row remains unchanged. This process is repeated until $K$ users have been assigned.

Note that since every row is strictly decreasing, selecting a later gain from a row is never beneficial before selecting all preceding gains from the same row. Consequently, the greedy procedure always selects the globally largest feasible gain at each iteration and solves \eqref{eq:max_gain} exactly.
\qed

\section*{Acknowledgments}
The authors would like to acknowledge Miguel Ángel Lagunas for the valuable discussions that helped improve this work.

\bibliographystyle{IEEEbib}
\bibliography{refs}

@ARTICLE{ota_original,
  author={Nazer, Bobak and Gastpar, Michael},
  journal={IEEE Transactions on Information Theory}, 
  title={Computation Over Multiple-Access Channels}, 
  year={2007},
  volume={53},
  number={10},
  pages={3498-3516},
  doi={10.1109/TIT.2007.904785}}

@ARTICLE{sp_magazine,
  author={Pérez-Neira, Ana and Martinez-Gost, Marc and Şahin, Alphan and Razavikia, Saeed and Fischione, Carlo and Huang, Kaibin},
  journal={IEEE Signal Processing Magazine}, 
  title={Waveforms for Computing Over the Air: A groundbreaking approach that redefines data aggregation}, 
  year={2025},
  volume={42},
  number={2},
  pages={57-77},
  doi={10.1109/MSP.2024.3500775}}

@ARTICLE{tbma_tong,
  author={Mergen, G. and Tong, L.},
  journal={IEEE Transactions on Signal Processing}, 
  title={Type based estimation over multiaccess channels}, 
  year={2006},
  volume={54},
  number={2},
  pages={613-626},
  doi={10.1109/TSP.2005.861896}}

@inproceedings{sahin_fsk,
  title={Distributed learning over a wireless network with FSK-based majority vote},
  author={{\c{S}}ahin, Alphan and Everette, Bryson and Hoque, Safi Shams Muhtasimul},
  booktitle={2021 4th International Conference on Advanced Communication Technologies and Networking (CommNet)},
  pages={1--9},
  year={2021},
  organization={IEEE}
}

@INPROCEEDINGS{tbma_fsk,
  author={Martinez-Gost, Marc and Pérez-Neira, Ana and Lagunas, Miguel Ángel},
  booktitle={GLOBECOM 2023 - 2023 IEEE Global Communications Conference}, 
  title={Frequency Modulation Aggregation for Federated Learning}, 
  year={2023},
  volume={},
  number={},
  pages={1878-1883},
  doi={10.1109/GLOBECOM54140.2023.10437413}}

@INPROCEEDINGS{tbma_robust,
  author={Martinez-Gost, Marc and Pérez-Neira, Ana and Lagunas, Miguel Ángel},
  booktitle={2025 33rd European Signal Processing Conference (EUSIPCO)}, 
  title={Robust Over-the-Air Computation with Type-Based Multiple Access}, 
  year={2025},
  volume={},
  number={},
  pages={2042-2046},
  doi={10.23919/EUSIPCO63237.2025.11226299}}

@article{sahin2026generic,
  title={A Generic Multi-dimensional Symbol Construction for Digital Over-the-Air Computation and Practical Aspects},
  author={Sahin, Alphan},
  journal={arXiv preprint arXiv:2606.18085},
  year={2026}
}

@inproceedings{sahin_sync,
  title={On the feasibility of distributed phase synchronization for coherent signal superposition},
  author={{\c{S}}ahin, Alphan},
  booktitle={2025 IEEE 36th International Symposium on Personal, Indoor and Mobile Radio Communications (PIMRC)},
  pages={1--6},
  year={2025},
  organization={IEEE}
}

@ARTICLE{power_control,
  author={Cao, Xiaowen and Zhu, Guangxu and Xu, Jie and Huang, Kaibin},
  journal={IEEE Transactions on Wireless Communications}, 
  title={Optimized Power Control for Over-the-Air Computation in Fading Channels}, 
  year={2020},
  volume={19},
  number={11},
  pages={7498-7513},
  doi={10.1109/TWC.2020.3012287}}

@book{verdu1998multiuser,
  title={Multiuser detection},
  author={Verdu, Sergio},
  year={1998},
  publisher={Cambridge university press}
}

@inproceedings{martinez2023lora,
  title={LoRa-based over-the-air computing for Sat-IoT},
  author={Martinez-Gost, Marc and P{\'e}rez-Neira, Ana and Lagunas, Miguel Angel},
  booktitle={2023 31st European Signal Processing Conference (EUSIPCO)},
  pages={1514--1518},
  year={2023},
  organization={IEEE}
}

@article{spectrum_apn,
  author    = {Perez, Ana and Sahin, Alphan},
  title     = {Teaching Radio Waves to Compute},
  journal   = {IEEE Spectrum},
  volume    = {63},
  number    = {5},
  pages     = {24--31},
  year      = {2026},
  month     = {May},
  publisher = {IEEE}
}

@article{martinez2026tbma_ml,
  title={Low-Complexity Maximum Likelihood Detection for Type-Based Over-the-Air Computation},
  author={Martinez-Gost, Marc and Lagunas, Miguel {\'A}ngel and P{\'e}rez-Neira, Ana},
  journal={arXiv preprint arXiv:2609.10136},
  year={2026}
}

@ARTICLE{goldenbaum,
  author={Goldenbaum, Mario and Boche, Holger and Stańczak, Sławomir},
  journal={IEEE Transactions on Signal Processing}, 
  title={Harnessing Interference for Analog Function Computation in Wireless Sensor Networks}, 
  year={2013},
  volume={61},
  number={20},
  pages={4893-4906},
  doi={10.1109/TSP.2013.2272921}}

@ARTICLE{channelcomp,
  author={Razavikia, Saeed and Barros da Silva, José Mairton and Fischione, Carlo},
  journal={IEEE Transactions on Communications}, 
  title={ChannelComp: A General Method for Computation by Communications}, 
  year={2024},
  volume={72},
  number={2},
  pages={692-706},
  doi={10.1109/TCOMM.2023.3324999}}

@inproceedings{mapreduce,
 author = {Chu, Cheng-tao and Kim, Sang and Lin, Yi-an and Yu, Yuanyuan and Bradski, Gary and Olukotun, Kunle and Ng, Andrew},
 booktitle = {Advances in Neural Information Processing Systems},
 editor = {B. Sch\"{o}lkopf and J. Platt and T. Hoffman},
 pages = {},
 publisher = {MIT Press},
 title = {Map-Reduce for Machine Learning on Multicore},
 url = {https://proceedings.neurips.cc/paper_files/paper/2006/file/77ee3bc58ce560b86c2b59363281e914-Paper.pdf},
 volume = {19},
 year = {2006}
}

@INPROCEEDINGS{consensus,
  author={Lee, Jihoon and Jang, Yonghoon and Kim, Hansol and Kim, Seong-Lyun and Ko, Seung-Woo},
  booktitle={ICC 2023 - IEEE International Conference on Communications}, 
  title={Over-the-Air Consensus for Distributed Vehicle Platooning Control}, 
  year={2023},
  volume={},
  number={},
  pages={5965-5971},
  doi={10.1109/ICC45041.2023.10279484}}

@ARTICLE{robust_sync,
  author={Goldenbaum, Mario and Stanczak, Slawomir},
  journal={IEEE Transactions on Communications}, 
  title={Robust Analog Function Computation via Wireless Multiple-Access Channels}, 
  year={2013},
  volume={61},
  number={9},
  pages={3863-3877},
  doi={10.1109/TCOMM.2013.072913.120815}}

@ARTICLE{oac_mimo,
  author={Zhu, Guangxu and Huang, Kaibin},
  journal={IEEE Internet of Things Journal}, 
  title={MIMO Over-the-Air Computation for High-Mobility Multimodal Sensing}, 
  year={2019},
  volume={6},
  number={4},
  pages={6089-6103},
  doi={10.1109/JIOT.2018.2871070}}

@ARTICLE{survey,
  author={Şahin, Alphan and Yang, Rui},
  journal={IEEE Communications Surveys \& Tutorials}, 
  title={A Survey on Over-the-Air Computation}, 
  year={2023},
  volume={25},
  number={3},
  pages={1877-1908},
  doi={10.1109/COMST.2023.3264649}}

@book{proakis2008digital,
  title={Digital Communications},
  author={Proakis, J.G. and Salehi, M.},
  isbn={9780071263788},
  lccn={2007036509},
  series={McGraw-Hill International Edition},
  url={https://books.google.es/books?id=ksh0GgAACAAJ},
  year={2008},
  publisher={McGraw-Hill}
}

@ARTICLE{emergency,
  author={Feng, Zebiao and Huang, Weidong and Cao, Jie},
  journal={IEEE Communications Magazine}, 
  title={Active Outage Mapping for Emergency Communications with Uncertainty-Aware Radio Environment Maps}, 
  year={2026},
  volume={64},
  number={9},
  pages={46-52},
  doi={10.1109/MCOM.001.2500697}}

@ARTICLE{em_processing,
  author={Dardari, Davide and Torcolacci, Giulia and Pasolini, Gianni and Decarli, Nicolò},
  journal={IEEE Signal Processing Magazine}, 
  title={Over-the-Air Electromagnetic Signal Processing: From fundamentals to emerging paradigms}, 
  year={2026},
  volume={43},
  number={1},
  pages={6-28},
  doi={10.1109/MSP.2025.3622001}}

@article{dahl2026decentralized,
  title={Decentralized Power Control for Over-the-Air Computation with Phase Noise},
  author={Dahl, Martin and Larsson, Erik G},
  journal={arXiv preprint arXiv:2606.19102},
  year={2026}
}

@ARTICLE{index_modulation,
  author={Basar, Ertugrul},
  journal={IEEE Communications Magazine}, 
  title={Index modulation techniques for 5G wireless networks}, 
  year={2016},
  volume={54},
  number={7},
  pages={168-175},
  doi={10.1109/MCOM.2016.7509396}}

@ARTICLE{log-fsk,
  author={Martinez-Gost, Marc and Pérez-Neira, Ana and Ángel Lagunas, Miguel},
  journal={IEEE Transactions on Communications}, 
  title={Expanding Over-the-Air Computation With Frequency Modulations}, 
  year={2025},
  volume={73},
  number={11},
  pages={12430-12440},
  doi={10.1109/TCOMM.2025.3576925}}

@INPROCEEDINGS{tbma-unsourced,
  author={Ngo, Khac-Hoang and Krishnan, Deekshith Pathayappilly and Okumus, Kaan and Durisi, Giuseppe and Ström, Erik G.},
  booktitle={2024 IEEE 25th International Workshop on Signal Processing Advances in Wireless Communications (SPAWC)}, 
  title={Type-Based Unsourced Multiple Access}, 
  year={2024},
  volume={},
  number={},
  pages={911-915},
  doi={10.1109/SPAWC60668.2024.10694658}}

@INPROCEEDINGS{ngo-feel,
  author={Okumus, Kaan and Ngo, Khac-Hoang and Ganesan, Unnikrishnan Kunnath and Durisi, Giuseppe and Ström, Erik G. and Pandey, Shashi Raj},
  booktitle={ICC 2026 - IEEE International Conference on Communications}, 
  title={Type-Based Unsourced Federated Learning With Client Self-Selection}, 
  year={2026},
  volume={},
  number={},
  pages={1-6},
  doi={10.1109/ICC59461.2026.11587510}}

\end{document}